\documentclass[a4paper]{jpconf}
\usepackage{graphicx,mathrsfs}
\newcommand{\be}{\begin{equation}}
\newcommand{\ee}{\end{equation}}
\newcommand{\bea}{\begin{eqnarray}}
\newcommand{\eea}{\end{eqnarray}}

\begin{document}
\title{Selected problems in astrophysics of compact objects }

\author{Armen Sedrakian}

\address{Institute for Theoretical Physics, 
J.~W.~Goethe-University, \\D-60438  Frankfurt-Main, Germany
}

\ead{sedrakian@th.physik.uni-frankfurt.de}

\begin{abstract}
  I review three problems in astrophysics of compacts stars: (i) the
  phase diagram of warm pair-correlated nuclear matter at
  sub-saturation densities at finite isospin asymmtery; (ii) the
  Standard Model neutrino emission from superfluid phases in neutron
  stars within the Landau theory of Fermi (superfluid) liquids; (iii)
  the beyond Standard Model physics of axionic cooling of compact
  stars by the Cooper pair-breaking processes.
\end{abstract}

\section{Introduction}

The present lecture summarizes some of the recent work on three
problems in the astrophysics of compact stars listed in the
abstract. It attempts to provide sufficient perspective and detail
within the limited span of pages as to be a useful introduction to
these problems. The selection of the topics is largely motivated by
the author's interests and covers only a small part of many facets of
compact star physics.

Neutron stars are born in spectacular explosions of type-II
supernovas~\cite{Burrows:2012ew}.  The equation of state of matter at
subnuclear densities and more generally its phase structure and
composition are important ingredients of supernova physics, one key
reason being their importance for the formation of neutrino signal at
the neutrino-sphere and for transport of the energy that ultimately
may generate a successful explosion.  The temperature range relevant
for subnuclear densities is several MeVs. In Sec.~\ref{sec:2} we will
discuss some recent progress in the understanding of the phase diagram of
nuclear matter in this regime.

If a neutron star is formed in a supernova explosion, its further
thermal evolution is determined by the neutrino emission from the interior of
the star: neutrinos once produced in the star escape the stellar
environment without further interactions
\cite{Yakovlev:2007vs,Page:2012,Sedrakian:2006mq}.  The black-body
radiation form the surface of the star in thermal equilibrium, while
providing us with the information on the composition of matter and
effective temperature of the emitting regions, is an unimportant
cooling agent until the star is hundreds of thousands years old. The
physics of neutrino emission from the interiors of neutron stars is
discussed in Sec.~\ref{sec:3} in the case of (non-exotic) baryonic
matter.

Non-Standard Model particles emitted during the neutrino stage of
cooling of a neutron star can drain sufficient energy as to ``spoil''
the Standard Model based cooling scenario. This may be used to place
limits on the properties of exotic particles, such as the
axions. Astrophysical bounds on the properties of axions, i.e., the
mass and the coupling to the Standard Model particles, have been
derived from stellar (red giant, white dwarf, supernova, and neutron
star) physics, which are complementary to those obtained from
cosmology and terrestrial laboratory
searches~\cite{Raffelt:2006cw}. Section~\ref{sec:4} discusses the
bounds on the axion properties derived from the cooling of neutron
stars through inclusion of a new process of axion emission via the
Cooper pair-breaking in superfluid baryonic matter~\cite{Keller:2012yr}.

\section{Warm isospin asymmetrical nuclear matter}
\label{sec:2}

Nuclear matter at sub-saturation densities, i.e., at $\rho \le
0.5\,\rho_0$, where $\rho_0 = 2.8\times 10^{14}$g cm$^{-3}$ is the
nuclear saturation density, is a many-body system with well defined
pair interactions, whose dominant attractive part is responsible for
the formation of nuclear clusters and the Bardeen-Cooper-Schrieffer
(BCS) type pair-condensate(s) of nucleons. In the supernova context
the matter is at finite, but small compared to cold neutron stars,
isospin asymmetry. Furthermore, low densities and relatively high
(again compared to neutron stars) temperatures allow for the existence
of a substantial amount of light clusters, notably deuterons, tritons,
He$^3$ nuclei and alpha
particles~\cite{Sumiyoshi:2008qv,Heckel:2009br,Typel:2009sy,Hempel:2009mc,Raduta:2010ym,Oertel:2012qd,Gulminelli:2011hr}.

Because the unbound nucleons and deuterons are the dominant component of
the matter at all relevant densities it is useful to consider only
two-body correlations first. In the extreme low density limit such
matter can be considered as a mixture of quasi-free nucleons and
deuterons, where the only effect of the interaction is to renormalize
the mass of the constituents. The deuterons being bosons may also
condense and from  a Bose-Einstein condensate (BEC) even without
interactions if, of course, the temperature is sufficiently
low~\cite{Alm:1993zz,Baldo:1995zz,Sedrakian:2005db,Mao:2008wz,Huang:2010fk,Jin:2010nj}.
Imagine now increasing density while keeping the temperature of the
system constant, such that the degeneracy of the system is effectively
increased. As the density $\rho\to \rho_0$ the abundances of deuterons
are reduced because of the Pauli-blocking of the phase space
available to nucleons~\cite{Sedrakian:2005db,Ropke:2012qv}.  However,
the emergence of the Fermi-surface of nucleons (at finite isospin
- two Fermi surfaces of neutrons and of protons) leads to  BCS type
correlations and formation of macroscopic coherent state - the
condensate of nucleons. The interaction channel responsible for this
pairing phenomenon is the $^3S_1$-$^3D_1$ partial wave, i.e., the same
interaction channel which binds the deuteron. Thus, it has been
conjectured that the nuclear matter may undergo a BCS-BEC phase
transition, first in the context of intermediate energy heavy-ion
collisions~\cite{Alm:1993zz,Baldo:1995zz} and more recently in the
context of supernovas~\cite{Heckel:2009br,Stein:2012wd}.

The theoretical framework for the description of the BCS-BEC
transition was developed by Nozi\`eres and Schmitt-Rink in condensed
matter physics~\cite{Nozieres:1985zz}.  Isospin asymmetry, induced by
weak interactions in stellar environments and expected in exotic
nuclei, disrupts isoscalar neutron-proton ($np$) pairing, since the
mismatch in the Fermi surfaces of protons and neutrons suppresses the
pairing correlations~\cite{Sedrakian:1999cu}.  The standard
Nozi\`eres-Schmitt-Rink theory of the BCS-BEC crossover must also be
modified, such that the low-density asymptotic state becomes a gaseous
mixture of neutrons and deuterons~\cite{Lombardo:2001ek}.

Two relevant energy scales for the problem domain under study are
provided by the shift $\delta\mu = (\mu_n - \mu_p)/2$ in the chemical
potentials $\mu_n$ and $\mu_p$ of neurons and protons from their
common value $\mu_0$ and the pairing gap $\Delta_0$ in the $SD$
channel at $\delta\mu=0$.  With increasing isospin symmetry, i.e., as
$\delta\mu$ increases from zero to values of order $\Delta_0$, a
sequence of unconventional phases may emerge.  One of these is a
neutron-proton condensate whose Cooper pairs have non-zero
center-of-mass (CM) momentum~\cite{Sedrakian:2000an,Muther:2002dm}.
This phase is the analogue of the Larkin-Ovchinnikov-Fulde-Ferrell
(LOFF) phase in electronic superconductors~\cite{LO,FF}. Another
possibility is phase separation into superconducting and normal
components, proposed in the context of cold atomic
gases~\cite{Bedaque:2003hi}.  At large isospin asymmetry, where $SD$
pairing is strongly suppressed, a BCS-BEC crossover may also occur in
the isotriplet $^1S_0$ pairing channel.  The ideas of unconventional
$SD$ pairing and the BCS-BEC crossover in a model of
isospin-asymmetric nuclear matter where combined recently in
Ref.~\cite{Stein:2012wd}. A phase diagram for superfluid nuclear
matter over wide ranges of density, temperature, and isospin asymmetry
was constructed.  A self-consistent set of equations, which includes
the gap equation and the expressions for the densities of constituents
(neutrons and protons) was solved allowing for the ordinary BCS state,
its low-density asymptotic counterpart BEC state and two phases that
owe their existence to the isospin asymmetry: the phase with a moving
condensate (LOFF phase) and the phase where the normal fluid and
superfluid break down into separate domains.

\begin{figure}[tb]
\begin{center}
\includegraphics[width=11.5cm,height=9cm]{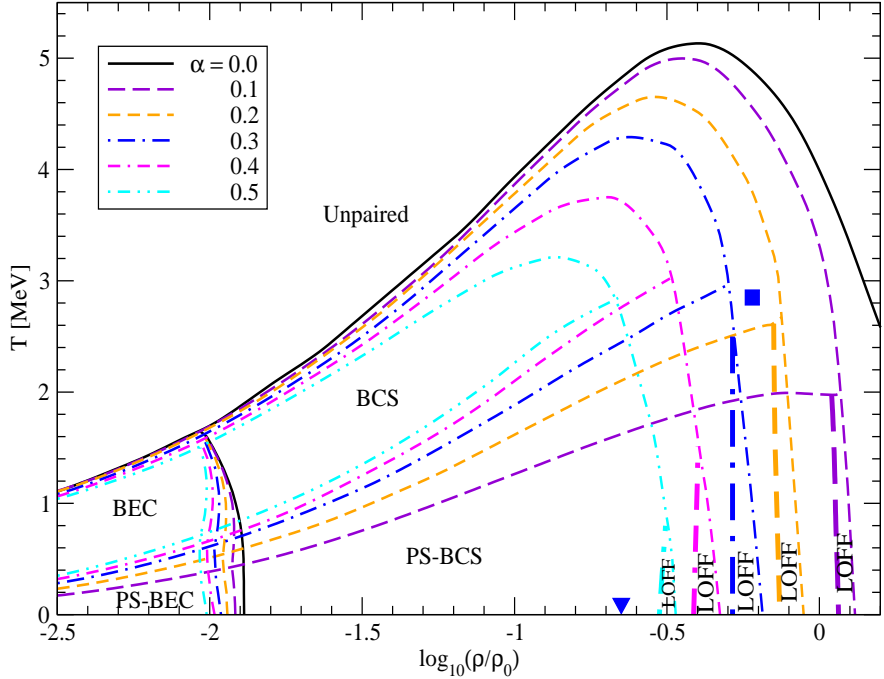}
\caption{Phase diagram of dilute nuclear matter in the
  temperature-density plane for several isospin asymmetries $\alpha$
  (from Ref.~\cite{Stein:2012wd}).  Included are four phases: unpaired
  phase, BCS (BEC) phase, LOFF phase, and PS-BCS (PS-BEC) phase. For
  each asymmetry there are two tri-critical points, one of which is
  always a Lifshitz point~\cite{Chaikin}. For special values of
  asymmetry these two points degenerate into single tetra-critical
  point at $\log(\rho/\rho_0) = -0.22$ and $T = 2.85$ MeV (shown by a
  square dot) for $\alpha_4 = 0.255$. The LOFF phase disappears at the
  point $\log(\rho/\rho_0) = -0.65$ and $T=0$ (shown by a triangle)
  for $\alpha = 0.62.$ }
\label{fig:2}
\end{center}
\end{figure}

The phase diagram found in Ref.~\cite{Stein:2012wd} is shown in Fig~\ref{fig:2}.
The results for the BCS phase and BCS-BEC
crossover are consistent with the earlier studies: one observes a smooth
crossover to an asymptotic state corresponding to a mixture of a
deuteron Bose condensate and a gas of excess neutrons. The transition
from BCS to BEC is established according the following criteria:
(i)~The average chemical potential $\bar \mu$ changes its sign from
positive to negative values, (ii)~the coherence length of a Cooper
pair becomes comparable to the interparticle distance, i.e., $\xi \sim
d\sim \rho^{-1/3}$ as conditions change from $\xi \gg d$ to $\xi \ll
d$.

The nuclear LOFF phase arises as a result of the energetic advantage
of translational symmetry breaking by the condensate, in which pairs
acquire a non-zero CM momentum $\vec Q$. At constant asymmetry, a
temperature increase shifts the gap maximum and the free-energy
minimum of the LOFF phase toward small $Q$, and at sufficiently high
temperature and small asymmetry the BCS state is favored over the LOFF
phase.  This behavior is well understood in terms of the phase-space
overlap of the Fermi surfaces of neutrons and protons, which (at
finite asymmetry) increases with temperature and the momentum $Q$ of
the Cooper pairs.

Thus, as the temperature increases, we expect a restoration of the BCS
phase and of the translational symmetry in the superfluid.  Obviously,
the same restoration occurs when the isospin asymmetry is small
enough.  The superfluid phase with phase separation (PS) has the
symmetrical BCS phase as one of its components. The temperature
dependence of this phase is well established within BCS theory. The
second component, which accommodates the neutron excess, is a normal
Fermi liquid whose low-temperature thermodynamics is controlled by the
excitations in the narrow strip of width $T/\epsilon_{F,n/p}$ around
the Fermi surfaces of neutrons and protons.

The transition to the BEC regime of strongly-coupled neutron-proton
pairs, which are asymptotically identical with deuterons, occurs at
low densities. As already well established, in the case of neutron-proton
pairing the criteria for the BCS-BEC transition are fulfilled, i.e.,
$\bar \mu$ changes sign and the mean distance between the pairs 
becomes larger than the coherence length of the superfluid.

We now turn to the question of how the BCS-BEC crossover is affected
by the existence of nuclear LOFF and PS phases at non-zero isospin
asymmetries, and conversely how these phases evolve in the
strongly-coupled regime if the density of the system is decreased.
Four different phases of matter are present in the phase
diagram (Fig.~\ref{fig:2}).  (i) The unpaired phase is always the
ground state of matter at sufficiently high temperatures $T>T_{c0}$,
where $T_{c0}(\rho)$ is the critical temperature of the superfluid
phase transition at $\alpha =0$. (ii) The LOFF phase is the ground
state in a narrow temperature-density strip at low temperatures and
high densities. (iii) The PS phase appears at low temperatures and low
densities, while the isospin-asymmetric BCS phase is the ground state
for all densities at intermediate temperatures. In the extreme
low-density and strong-coupling regime the BCS superfluid phases have
two counterparts: the BCS phase evolves into the BEC phase of
deuterons, whereas the PS-BCS phase evolves into the PS-BEC phase, in
which the superfluid fraction of matter is a BEC of deuterons.  The
superfluid-unpaired phase transitions and the phase transitions
between the superfluid phases are of second order (thin solid lines in
Fig.~\ref{fig:2}), with the exception of the PS-BCS to LOFF
transition, which is of first order (thick solid lines in
Fig.~\ref{fig:2}). The BCS-BEC transition and the PS-BCS to PS-BEC
transition are smooth crossovers.  At non-zero isospin asymmetry the
phase diagram features two tri-critical points where the simpler
pairwise phase coexistence terminates and three different phases
coexist. 

The extreme low-density region of the phase diagram features two
crossovers. At intermediate temperatures one recovers the well-known
BCS-BEC crossover, where the neutron-proton BCS condensate transforms
smoothly into a BEC gas of deuterons with some excess of neutrons.
The new ingredient of our phase diagram is the second crossover at low
temperatures, where the heterogeneous superfluid phase is replaced by
a heterogeneous mixture of a phase containing a deuteron condensate and
a phase containing neutron-rich unpaired nuclear matter. 

What would be the effect of adding somewhat heavier clusters, notably
$^3$He, $^3$H and $^4$He, to the phases discussed above? At low
densities statistical equilibrium suggests that the heavier clusters
may ``eat up'' some of the phase-space, therefore their main effect
would be to reduce the phase space available to pair-correlated
particles, which would eventually lead to some shifts in the phase
diagram without qualitative modifications of its structure or
topology. The physics might not be as trivial at high densities, for
we know that the asymptotically dense phase of matter does not contain
clusters. The transition to the homogenous phase can be attributed to
the Pauli-blocking of the phase space that can be used to form three
and four body bound states. In other words, the rise of the Fermi
surface suppresses the formation of three and four-body bound states.
It remains an open problem to tackle the interplay of the clusters and
pair-correlations on the ultimate composition of matter at subnuclear
densities. As stated in the introduction, the answer is not merely of
academic interest. The neutrinos (and more generally leptons)
transport energy in the supernova processes while moving through this
environment; its composition, many-body effects, etc remain the key
unknowns in the accurate description of the neutrino transport.

\section{Neutrino pair-breaking processes}
\label{sec:3}

Next we trasport ourselves in time several weeks past the supernova
explosion and the formation of a neutron star. During this period of
time the star has cooled down to the temperature of superfluid phase
transition of baryons, which is roughly in the range $T_c\simeq 0.5$
MeV. At this stage the crust of the star cools predominantly via the
electron ($e^-$) bremsstrahlung process: $e^-+A\to e^-+A+\nu+\bar
\nu$, where $A$ stands for a nucleus and $\nu$ and $\bar \nu$ for the
neutrino and antineutrino~\cite{Shapiro}.  The surface of the star
cools by emitting thermal soft $X$-rays (photons). Both processes are
not important sinks of energy up until the star's age $t\ge 10^4-10^5$
yr~\cite{Yakovlev:2007vs,Page:2012,Sedrakian:2006mq}.  The exact value
of the transition to the crust plus surface cooling depends on a
number of factors, which we will not discuss. While it was widely
accepted from the beginning that the superfluidity will suppress the
neutrino radiation processes on baryons in the core, once it becomes
superfluid, the less trivial pair-breaking processes where initially
neglected, although the rates of these processes, computed at
one-loop, where available since
1976~\cite{Flowers:1976ux,Voskresensky:1987hm,Kaminker:1999ez}.  The
Standard Model weak neutral interactions proceed via vector and axial
vector interactions. Thus, the neutrino emissivity (phase space
integrated rate of neutrino emission) is mainly determined by the
response functions of matter to weak vector and axial-vector
currents. These can be computed within the finite temperature Green's
functions technique in general. Specific applications within the
real-time Schwinger-Keldysh formalism can be found in
Refs.~\cite{Voskresensky:1987hm,Sedrakian:1999jh}.

Before discussing these response functions, we pause to recall the
many-body developments of the formal theory.  The polarization tensor
in superfluid system involves a re-summation of infinite series of
particle-hole diagrams. The methods for doing so were developed in the
case of symmetrical nuclear matter, with the purpose of applications
to finite nuclei in Ref.~\cite{Migdal:1967}.  This work is one of the
first applications of the Landau Fermi liquid theory (which was
designed to describe the properties of non-superfluid liquid He$^3$)
to superfluid systems. The diagrammatic methods of re-summation of
particle-hole ladders were developed even
earlier~\cite{AG,Abrikosov:1962}.  While the theory of superfluid
Fermi liquids was generalized further by Leggett \cite{Leggett:1966zz}
to finite temperatures (in the context of condensed matter theories),
Leggett did so by computing directly the response functions and not
the vertex functions.  In nuclear physics, the theory took a turn that
focused the entire subsequent work on finite systems (i.e. nuclei),
which were treated in the equation of motion formalism for
second-quantized operators.

The importance of the vertex corrections for the case of the vector
current interactions was first pointed out in
Ref.~\cite{Leinson:2006}, where the authors implemented a polarization
tensor adopted from condensed matter work (derived not quite in the
regime needed for neutron stars). A number of subsequent works derived
directly the finite temperature vertex functions and polarization
tensors for neutron and proton components of the core of the star
\cite{Sedrakian:2006ys,Kolomeitsev:2008mc,Steiner:2008qz,Sedrakian:2012ha}. 
The current consensus is that the vector current emissivity is suppressed
compared to the one-loop results by a factor $(v_f/c)^4$, where
$v_f/c\sim 0.1$ is the Fermi velocity of the baryons in units of the
speed of light. At the same time it was established that the axial
vector interactions are adequately represented by the bare vertices
and one-loop polarization tensors. Explicitly, the axial vector
emissivity given by~\cite{Kaminker:1999ez,Kolomeitsev:2008mc}
\be \epsilon_{\nu}=\frac{24G_F^2g_A^2}{105\pi^3}
\nu(0) {v_f}^2T^7 I_{\nu}, \quad I_{\nu} = z^7 \int_1^{\infty} dy
\frac{y^5}{\sqrt{y^2-1}} f_F \left(z y\right)^2 , \ee where 
 $G_F = 1.166\times 10^{-5}$ GeV$^{-2}$ is 
the Fermi weak coupling constant, $g_A = 1.25$ is (unquenched) axial
vector coupling constant, $\nu(0)$ is the density of states at the
Fermi surface, $T$ is the temperature, $z= \Delta(T)/T$ is the ratio
of the pairing gap to the temperature, and $f_F (x) =
[1+\exp(x)]^{-1}$ is the Fermi distribution function.  The emissivity
of the axial vector current is of the order $(v_f/c)^2$ and,
therefore, is the dominant channel of the energy loss.

\section{Axion emission from compact stars}
\label{sec:4}
The strong sector of the Standard Model may feature a CP violating
interaction, which arises due to a topological interaction term in the
QCD Lagrangian~\cite{'tHooft:1976up}
\be
\label{thetaaction}
\mathscr{L}_\theta={g^2\theta\over 32\pi^2}
 \,F^a_{\mu\nu}\tilde F^{\mu\nu a}, 
\ee
where $F_{\mu\nu}^a
= \partial_{\mu}A_{\nu}-\partial_{\nu}A_{\mu}+gf^{abc}A_{\mu b} A_{\nu
  c}$ is the gluon field strength tensor, $\tilde F^a_{\mu\nu} =
\epsilon_{\mu\nu\lambda\rho} F^{\lambda\rho a}/2$, $f^{abc}$ are the
structure constants of $SU(3)$ group, $\theta$ is the parameter which
parametrizes the non-perturbative vacuum states of QCD
$\vert\theta\rangle = \sum_n \exp(-in\theta)\vert n\rangle$, where $n$
is the winding number characterizing each distinct state of QCD, which
is not connected to another by any gauge
transformation.  The QCD action changes by
$2\pi$ under the shift $\theta\to\theta+2\pi$, i.e., $\theta$ is a
periodic function with a period of $2\pi.$ In presence of quarks the
physical parameter is not $\theta$, but
$
\bar\theta = \theta + \arg\det m_q , 
$ where $m_q$ is the matrix of
quark masses. Experimentally, the upper bound on the value of this
parameter is $\bar\theta\le 10^{-10}$, which is based on the
measurements of the electric dipole moment of neutron $d_n<6.3\cdot
10^{-26}e$ cm~\cite{Baker:2006ts}.  The smallness of $\bar\theta$ is
the strong CP problem:  the Standard Model does not provide any
explanation on why this number should not be of order unity.

An elegant solution to the strong CP problem is provided by the
Peccei-Quinn
mechanism~\cite{Peccei:1977hh,Weinberg:1977ma,Wilczek:1977pj}.  This
solution amounts to introducing a global $U(1)_{PQ}$ symmetry, which
adds an additional anomaly term to the QCD action proportional to the
axion field $a$. This term acts as a potential for the axion field and
gives rise to an expectation value of the axion field $\langle a
\rangle \sim -\bar\theta$.  The physical axion field is then $a-
\langle a \rangle $, so that the undesirable $\theta$ term in the
action is replaced by the physical axion field. The axion is the
Nambu-Goldstone boson of the Peccei-Quinn $U(1)_{PQ}$ symmetry
breaking~\cite{Weinberg:1977ma,Wilczek:1977pj},
 and its effective Lagrangian has the form \be \mathscr{L}_a =
-\frac{1}{2}\partial_{\mu} a\partial^{\mu} a +
\mathscr{L}_{int}(\partial_{\mu} a,\psi), \ee where the second term
describes the coupling of the axion to fermion fields ($\psi$) of the
Standard Model.

There are ongoing experimental searches for the axion.
Cosmology and astrophysics provide strong complementary constraints.
Because axions can be effectively produced in the interiors of stars
they act as an additional sink of energy. The requirements that the
energy loss from a star is consistent with the astrophysical
observations place lower bounds on the coupling of axions to the
standard model particles, and hence on the Peccei-Quinn symmetry
breaking scale~\cite{Raffelt:2006cw,Raffelt:2011ft}. The latter limit
translates into an upper limit on the axion mass.  Such arguments have
been applied to the physics of supernova explosions
~\cite{Brinkmann:1988vi,Burrows:1988ah,Raffelt:1993ix,Janka:1995ir,Hanhart:2000ae}
and white dwarfs~\cite{Corsico:2012ki}. In the case of supernova
explosions the dominant energy loss process is the emission of an
axion in the nucleon ($n$) bremsstrahlung $n+n\to n+n+a$. The same
process was considered earlier by Iwamoto as a cooling mechanism for
mature neutron stars, i.e., neutron stars with core temperature in the
range $10^8-10^9$~K~\cite{Iwamoto:1984ir}.  The implications of 
the axion emission by the modified Urca and nucleon bremsstrahlung, 
as calculated in Ref.~\cite{Iwamoto:1984ir}, were briefly studied 
in Ref.~\cite{Umeda:1997da}.
However, as discussed in the previous section the pair-breaking
processes are the dominant energy loss channels for temperatures below
the critical temperature. Therefore, to obtain bounds on the
properties of axions from the cooling of compact stars, the rate of
the axion cooling is required. This was computed recently in
Ref.~\cite{Keller:2012yr}  in the case of $S$-wave superfluid. 

\begin{figure*}[t]
\begin{center}
\includegraphics[height=1.2cm,width=12cm]{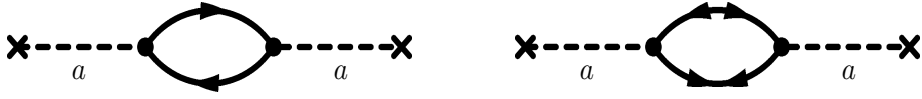}
\end{center}
\caption[] { The two diagrams contributing to the polarization tensor of
baryonic matter,  which defines  the axion emissivity. 
The ``normal'' baryon propagators for particles (holes) are shown 
by single-arrowed lines directed from left to right (right to left).
The double arrowed lines correspond to the ``anomalous'' 
propagators  with two incoming or outgoing
arrows.  The horizontal dashed lines represent the axion $a$.  }
\label{fig2}
\end{figure*}

We now briefly outline this calculation.  The energy radiated per unit
time in axions (axion emissivity) is given by the phase-space integral
over the probability of the emission process  \be\label{emiss1}
\epsilon_a = -f_a^{-2} \int\frac{d^3q}{(2\pi)^32\omega} ~\omega
g_B(\omega) \kappa_a, \quad \kappa_a= q_{\mu}q_{\nu} \Im
\Pi_a^{\mu\nu}(q), \ee where $q$ and $\omega$ are the axion momentum
and energy, $\Pi_a^{\mu\nu}(q)$ is the polarization tensor of the
superfluid baryonic matter, $g_B(\omega)$ is the Bose distribution
function, $f_a$ is the axion-baryon coupling strength.  The
polarization tensor of a superfluid obtains contributions from four
distinct diagrams that can be formed from the normal and anomalous
propagators with four distinct effective vertices~\cite{Sedrakian:2006ys,Kolomeitsev:2008mc,Steiner:2008qz,Sedrakian:2012ha}. However, for the
axial vector perturbations the vertices are not renormalized in the
medium and, therefore, one proceeds with the bare vertices, in which
case the number of the distinct contributions to the polarization
tensor reduces to a sum of two admissible bare loops (see
Fig.~\ref{fig2}).
The axion emissivity obtained from Eq. (\ref{emiss1}) is given by
\bea\label{eq:axion1} \epsilon_a &=& \frac{ 8 }{3\pi} \, f_a^{-2}
\,\nu(0)\, v_f^2 \, T^5 \, I_a, \quad I_a = z^5\int_1^{\infty}\!\! dy
~ \frac{y^3}{\sqrt{y^2-1}} f_F\left(z y\right)^2. \eea 
The $T^5$ scaling of the emissivity is
understood as follows. The integration over the phase space of
neutrons carries a power of $T$, since for degenerate neutrons the
phase-space integrals are confined to a narrow strip around the Fermi
surface of thickness $T$. The axion is emitted thermally and being
relativistic contributes a factor $T^3$ to the emissivity. The one
power of $T$ from the energy of the axion and the inverse one power of
$T$ from the energy conserving delta function cancel. The transition
matrix element is proportional to the combinations of $u_p$ and $v_p$
amplitudes, which are dimensionless, but contain {\it implicit}
temperature dependence due to the temperature dependence of the gap
function. This dependence is not manifest in Eq.~(\ref{eq:axion1}),
i.e., was absorbed in the definition of the integral $I_a$. Thus, the
explicit temperature dependence of the axion emission rate
Eq.~(\ref{eq:axion1}) is $T^5$.  As discussed in Sec.~\ref{sec:3} at
temperatures of order the critical temperature $T_c\simeq 10^9$~K the
superfluid cools primarily by emission of neutrinos via the
pair-breaking processes driven by the axial-vector currents (we
continue to assume that potential fast cooling via direct Urca
processes is prohibited).  
In a first approximation one may require that the axion luminosity
does not exceed the neutrino luminosity, i.e.,
\bea\label{eq:ratio}
\frac{\epsilon_a}{\epsilon_{\nu}} &=& 
\frac{10\pi^2}{f_a^2 G_F^2g_A^2\zeta_A }  \frac{I_a}{I_\nu}
 \simeq \frac{59.2}{ f_a^{2} G_F^2 \Delta(T)^2}~r(z)
< 1,
\eea 
where $r(z) \equiv z^2(I_a/I_{\nu}) $.  Not far from the critical
temperature $\Delta (T) \simeq 3.06T_c\sqrt{1-T/T_c}$, which
translates into $z = 3.06\, t^{-1}\sqrt{1-t}$, where $t =
T/T_c$. Numerical evaluations of the integrals provides the following
values $r(0.5) = 0.07$, $r(1) = 0.26$, $r(2) = 0.6$ and asymptotically
$r(z) \to 1$ for $z\gg 1$. Noting that $r(z)\le 1$ , one finally obtains
\be\label{eq:fbound} f_a > 5.92 \times 10^{9}\, \textrm{GeV} \,
\left[\frac {0.1~\textrm{MeV}}{\Delta(T)}\right],
\ee
which translates into an upper bound on the axion mass 
\be \label{eq:mbound} m_a = 0.62 \times
10^{-3}\, \textrm{eV}\, \left(\frac{10^{10}
    \textrm{GeV}}{f_a}\right)\le 1.05 \times 10^{-3}\, \textrm{eV}\,
\left[\frac {\Delta(T)}{0.1~\textrm{MeV}}\right].  
\ee 
The bound Eq.~(\ref{eq:fbound}) can be written in terms of the critical
temperature by noting that $\Delta (T) \simeq T_c$ in the temperature range $0.5\le
t< 1$ of interest.

The neutrino cooling era of compact stars, which spans the time-period
$ t\le 10^4-10^5$ yr after their birth in supernova explosions is a
sensitive probe of the particle physics of their interiors. If one
assumes that there are no rapid channels of cooling in neutron stars,
i.e., deconfined quarks, above Urca threshold fractions of protons or
hyperons (all of which lead to a rapid Urca cooling), then neutron
stars cool primarily by neutrino emission in Cooper pair-breaking
processes in baryonic superfluids. If, however, axions
exist in Nature, the neutron stars must cool via axion emission in
Cooper pair-breaking processes, whose axion emission rate scales as
$T^5$. This scaling differs from the $T^7$ scaling of the counterpart
neutrino processes. The difference arises from the
  different phase spaces required for the pairs of neutrinos and the
  axion and is independent of the baryonic polarization tensor. Note
  also that the rate of axion emission from a $P$-wave superfluid will
  differ from the $S$-wave rate, derived above, by a factor $O(1)$
  and, therefore, will not change quantitatively the obtained bounds
  on the axion parameters.  

  Similar bounds to those quoted above were obtained previously by
  Iwamoto~\cite{Iwamoto:1984ir} ($f /10^{10}\textrm{GeV} > 0.3$) from
  a comparison of the rates of axion bremsstrahlung and modified Urca
  neutrino emission by mature neutron stars, and by Umeda et
  al~\cite{Umeda:1997da} ($f /10^{10}\textrm{GeV} > 0.1-0.2$) from
  fits of cooling simulations to the PSR 0656+14 data\footnote{Note
    that Eq. (9) of this work is incorrect, which could be the reason
    why the limits on the axion's mass reported in their Figs. 2-4 are
    by an order of magnitude larger compared to those quoted in
    Ref.~\cite{Keller:2012yr}.}.  The lower bound on $f_a$ derived in
  Ref.~\cite{Keller:2012yr} is somewhat larger than the one that
  follows form the requirement that the axions do not ``drain'' too
  much energy from supernova process so that it
  fails~\cite{Raffelt:2006cw,Brinkmann:1988vi,Burrows:1988ah,Raffelt:1993ix,Janka:1995ir,Hanhart:2000ae}. Note
  the dependence of the bound Eq.~(\ref{eq:fbound}) on the pairing
  gap. This is an ingredient, which originates from superfluidity of cold neutron stars,
  that does not appear in other bounds on axion parameters.  Because,
  the magnitude and density dependence of the gap are not well-known
  (for a review see Ref.~\cite{Sedrakian:2006xm}) there remains enough
  room for speculations on the impact of the axions on the cooling of
  compact stars.

\section{Perspectives}

Although the equation of state and composition of the dilute and warm
nuclear matter have been studied for several decades, there remain a
number of unsettled issues related to the complex many-body character
of this system. The interplay of the pair-correlations and clustering
is one of the challenging problems. As we have seen the phase structure of
the matter at finite isospin becomes more complicated due to the
emergence of novel superconducting phases. It is very likely that
along with the deuteron condensate and its BCS counterpart a BEC of
alpha particles will coexist with these phases. Thus, one needs to
work out the physics of a mixtures of several superfluids phases in
this context.  How these new features will affect the neutrino
transport in supernovas remains an open question.

 Neutron star cooling offers a new playground for the studies of
 beyond Standard Model physics. The example presented above shows that
 with microscopic rates of the axion emission from superfluid phases
 at hand one can place useful bounds on the axion properties. Axions
 offer a unique channel of rapid energy loss by medium to low mass
 compact stars, where the central densities are below the direct Urca
 thresholds and/or below the density of deconfinement into quark
 matter. Thus, the long-standing paradigm that the light and
 medium mass neutron stars cannot cool rapidly will not hold should
 the axions exist in Nature and  couple to matter strong enough 
 to cool a compact star faster than the Standard Model neutrinos.

\section*{Acknowledgements}

I gratefully acknowledge the discussions and collaboration with
J. W. Clark, X.-G. Huang, J. Keller, and M. Stein on the topics
presented in this lecture.

\end{document}